\documentclass{ws-procs975x65}

\begin{document}



\title{FROM QUANTUM HYDRODYNAMICS TO QUANTUM GRAVITY
}

\author{GRIGORY VOLOVIK
}

\address{Low Temperature Laboratory,
Helsinki University of Technology\\
P.O.Box 2200, FIN-02015 HUT, Finland\\
and\\
L.D. Landau Institute for Theoretical Physics,
  Moscow 119334, Russia\\
\email{volovik@boojum.hut.fi}}








\begin{abstract}
We discuss some lessons from quantum hydrodynamics to quantum gravity.
\end{abstract}

\bodymatter

\section{Introduction}\label{intro}

In the presentations at the Session `Analog Models of and for General
Relativity' at 11 Marcel Grossmann Meeting, 
general relativity has been considered as emergent phenomenon.  General
approaches to emergent relativity have been analyzed 
\cite{Dreyer}. 
Particular example when gravity 
is induced in the low-energy corner of quantum condensed matter of the
proper universality class has been presented \cite{Book}. 
It was suggested that
induced metric for scalar field may lead to
superluminal propagation of scalar field and escape from the black hole
without violation of Lorentz invariance
\cite{Babichev}. On the kinematic level the metric field emerges in many
different systems, and this allows us to simulate (at least
theoretically) effects of relativistic quantum field theory (QFT) in
curved space. At the moment the most promising media for simulations are
Bose-Einstein condensate (BEC), where the propagation of phonons  is
identical to propagations of a massless scalar field on a curved
space-time. In particular, 
it was suggested  to use the renormalization techniques developed in
 QFT to study the depletion of BEC 
\cite{Fagnocchi}; in other presentation 
the stability of sonic horizons in BEC\cite{Cano}   
and the scattering problems on rotating acoustic black holes have been
discussed 
\cite{Cherubini}.
 Effective metric appears for light propagating in
non-linear dispersive dielectric media \cite{Klippert} and in moving 
media
\cite{Rosquist}; for surface waves -- ripplons -- propagating on the
surface of quantum liquids or at the interface between two superfluids
\cite{Horizons}. The latter allows us to study experimentally the
instability of the quantum vacuum in ergoregion.

Probably our experience with superfluids and BEC will give us some hints
for solution of the fundamental problems in gravity, such as quantum
gravity and gravitating vacuum energy. Here we shall discuss the quantum
hydrodynamics of BEC and superfluids. Both hydrodynamics and general
relativity are perfect classical theories. General relativity can be viewed  as the theory of  hydrodynamic type where the collective variables are the metric fields $g_{\mu\nu}$ \cite{Hu}. At the quantum level, quantum
hydrodynamics and quantum gravity also share many common features, e.g.
both have quadratic divergences.  This is the reason why the problem of
quantization of hydrodynamics is at least 65 years old (see quantization
of the macroscopic dynamics of liquid in the first Landau paper
\cite{LandauQH} on superfluidity of
$^4$He); it is almost as old as the problem of quantization of
gravity \cite{Bronstein}.  Thus the lessons from quantum
hydrodynamics could be useful for quantum gravity.

\section{Classical  hydrodynamics}
\subsection{Classical  hydrodynamics}

The first quantization scheme for hydrodynamics 
was suggested by Landau in 1941 when he developed the  theory of
superfluidity in liquid $^4$He
\cite{LandauQH}.  In
his approach Landau separated liquid $^4$He into two parts:  
the ground state (which we now call the vacuum) and 
quasiparticles -- excitations above the ground state (which we call
matter). Such separation into vacuum and matter is generic and is
applicable to relativistic quantum fields (RQF).  The Landau approach was
essentially  different from that of Tisza
\cite{Tisza}, who suggested to separate liquid $^4$He into the Bose
condensate and the non-condensed atoms. Tisza's approach does make
sense, especially for the dilute Bose gases, where the condensed fraction
can be easily detected. However, it is important that the dynamics of the
Bose condensate and the exchange of energy and atoms between the
condensed and non-condensed fractions, belong to high-energy microscopic
physics. On the other hand,  the low-energy behavior of the superfluid
liquids and gases is governed by the Landau hyrodynamics picture. In
particular, at zero temperature both condensed and non-condensed atoms
participate in the coherent motion of the quantum vacuum with the total
mass density $\rho$.  This is because at $T=0$ the whole liquid is in the
coherent state described by a single many-body wave function
\cite{Feynman}, and thus the whole liquid is involved in the superfluid
motion in agreement with Landau ideas.

According to Landau, the Hamiltonian of quantum hydrodynamics is
the classical energy of liquid where the classical fields, velocity 
${\bf v}$ and mass density $\rho$, are substituted by the corresponding
quantum operators $\hat{\bf v}$ and  $\hat\rho$. So let us start with the
classical hydrodynamic energy of the liquid:
\begin{equation}
H_{\rm hydro}(\rho,{\bf v})=\int d^3x\left(\frac{1}{2}\rho v^2+
\tilde\epsilon(\rho)\right)~,~\tilde\epsilon(\rho)=
\epsilon(\rho)-\mu\rho.
\label{ClassicalHydrodynamicH}
\end{equation} 
Here $\epsilon(\rho)$ is the energy of static liquid which only depends
on $\rho$: since we consider the vacuum of the liquid (i.e. without
excitations which will appear after quatization) it is assumed that the
temperature
$T=0$. We added here the term with Lagrangian multiplier --  the constant
chemical potential
$\mu$. This term does not change the hydrodynamic   equations, but it
allows us to study thermodynamics of the liquid.  For example, the
equilibrium mass density of static liquid is obtained by minimization of
the energy with taking into account the conservation of the total mass of
the liquid, which gives:
\begin{equation}
\frac{d\epsilon}{d\rho}=\mu~~,~~{\rm
or}~~\frac{d\tilde\epsilon}{d\rho}=0~ .
\label{EquilibriumLiquid}
\end{equation} 
The pressure of the liquid in  equilibrium at $T=0$  is
\begin{equation}
P =-\frac{d(V\epsilon(M/V))}{dV}=-\tilde\epsilon~,
\label{EquilibriumLiquid}
\end{equation}
where $M$ is the total mass of liquid. This suggests that the relation
between the pressure $P$ and energy $\tilde\epsilon$  can be considered
as the equation of state for vacuum, and this is true. Such equation of
state $P=-\tilde\epsilon$ is applicable to the ground state (vacuum) of
any system, relativistic or non-relativistic; it follows from the
general thermodynamic arguments and does not depend on the microscopic
physics of the vacuum state. It is also applicable to the vacuum of RQF.
So, further on we shall treat the quantities
$\epsilon_{\rm
vac}\equiv \tilde\epsilon$ and 
$P_{\rm vac}\equiv -\tilde\epsilon$ as vacuum energy density and vacuum
pressure correspondingly.

There is no
satisfactory  description of classical hydrodynamics in terms of
Lagrangian: this requires introduction of either artificial variables or
extra dimension. The hydrodynamic equations  can be obtained using the
Hamiltonian formalism of Poisson brackets. The Poisson brackets
between the hydrodynamic variables are universal, they are determined by
the symmetry of the system and do not depend on the Hamiltonian (cf.
\cite{DzyaloshinskiiVolovick}),  that is why it is not necessary to use
the microscopic quantum theory for their derivation. For classical
hydrodynamic variables
 ${\bf v}$ and $\rho$ one has the following Poisson brackets:
\cite{DzyaloshinskiiVolovick,PoissonVorticity}
\begin{eqnarray}
\left\{\rho({\bf r}_1),\rho({\bf r}_2)\right\}= 0~,
\label{PBDensityDensity}\\
\left\{{\bf v}({\bf r}_1),\rho({\bf r}_2)\right\}= -\nabla
\delta({\bf r}_1-{\bf r}_2)~,
\label{PBVelocityDensity}
\\
 \left\{ v_{i}({\bf r}_1),v_{j}({\bf r}_2)\right\}=-{1\over
\rho}e_{ijk}
 (\nabla\times {\bf
v})_k \delta( {\bf r}_1-{\bf r}_2)~.
\label{PoissonBracketsVelocity}
\end{eqnarray}
The same Poisson brackets are obtained from the commutation relations for
the corresponding quantum operators $\hat\rho$ and $\hat{\bf v}$ derived
by Landau \cite{LandauQH} which follow from microscopic physics. 
Using the Poisson brackets
(\ref{PBDensityDensity})-(\ref{PoissonBracketsVelocity}) and
the Hamiltonian in Eq.(\ref{ClassicalHydrodynamicH}), one obtains the
hydrodynamic equations:
\begin{eqnarray}
\partial_t \rho=\left\{H,\rho\right\}= -\nabla\cdot(\rho{\bf v})~,
\label{ContinuityrEquation}
\\
~\partial_t{\bf
v}=\left\{H,{\bf v}\right\}=-({\bf v}\cdot\nabla){\bf
v}-\nabla\frac{d\epsilon}{d\rho}~.
\label{EulerEquation}
\end{eqnarray}

There are no fundamental parameters in classical hydrodynamics. But 
there are dimensional variables which enter the classical
hydrodynamics: mass density
 $\rho$, and energy density $\epsilon(\rho)$; the speed of  sound $c$ is
$c^2=\rho(d^2\epsilon/d\rho^2)$. In principle, in liquids one can
construct the  ``fundamental''  parameters,  the values of $c=c_0$ and
$\rho=\rho_0$ under two conditions, when the liquid is: (i) static and in
equilibrium; and (ii) at zero external pressure. These two conditions give
$d\epsilon/d\rho|_{\rho_0}=\mu_0$ and $P=\mu_0\rho_0 - 
\epsilon(\rho_0)=-\tilde\epsilon(\rho)=0$ correspondingly. At zero
external pressure, i.e. in the absence of external environment, one
has
 \begin{equation}
\epsilon_{\rm vac}=-P_{\rm vac}=0~.
\label{Nullification}
\end{equation} 
The nullification of vacuum energy occurs for any
non-disturbed equilibrium vacuum.  

 \subsection{Vortex-free classical  hydrodynamics}
 
 If one is interested in the vortex-free flow only,  ${\bf
v}=\nabla\theta$,  the hydrodynamic equations for $\theta$ and $\rho$ can
be  obtained using the Lagrangian formalism. The corresponding
hydrodynamic Lagrangian is:  
 \begin{equation}
L_{\rm hydro}(\rho,\theta)=H_{\rm hydro}-\rho\partial_t\theta=
\frac{1}{2}\rho v^2+ \epsilon(\rho)-\rho\partial_t\theta~,~{\bf
v}=\nabla\theta~.
\label{ClassicalLagrangian}
\end{equation} 
The constant chemical potential $\mu$ is absorbed here by 
$\partial_t\theta$.
 
The Poisson brackets for the vortex degrees of freedom have  been
discussed in Refs. \cite{Rasetti,PoissonVorticity}. 

In linear approximation the Lagrangian (\ref{ClassicalLagrangian})
describes sound waves. Sound waves propagating over  background flow of
the inhomogeneous liquid can be obtained from the hydrodynamic equations
(\ref{ContinuityrEquation}) and (\ref{EulerEquation}); the rigorous
procedure can be found  in Refs.
\cite{Stone2000,Stone2002}. As
was first found by Unruh
\cite{Unruh} the flow of liquid has the same effect on propagation of
sound waves as the metric in general relativity on propagation of a
massless relativistic particle. The effective metric for sound waves
generated by ${\bf v}({\bf r},t)$ and $\rho({\bf r},t)$ is
\begin{equation}
 g_{00}=-{\rho\over  c}(c^2-{\bf v}^2) ~,~  g_{ij}=
{\rho\over c}\delta_{ij} ~,~
g_{0i}=-g_{ij}v^j ~,~
\sqrt{-g}={\rho^2\over c}~.
\label{CovarianAcousticMetric}
\end{equation}
This is the half of general relativity, since the effective metric obeys
the hydrodynamic equations rather than Einstein equations. However, this
is enough for simulations of aspects of general relativity
which do not depend on Eistein equations. For example, effects related to
behavior of quantum fields in curved space can be
reproduced \cite{Unruh,BLV}. 

The full general relativity can
be generated in fermionic vacua near the Fermi points 
\cite{FrogNielBook,Book,Horava}.  Fermi point is a generic singularity in
the Green's function which is protected by topology in momentum
space. Expansion near the Fermi point leads to chiral fermions, gauge
fields and gravity as effective fields in the low-energy corner. 
 
 \subsection{Extended classical hydrodynamics}
 
The most general classical hydrodynamics is obtained when one  introduces
corrections to classical hydrodynamics by adding the gradient
 terms. For static liquids and gases the important modification is the  
dependence of energy on the gradient of mass density:
 \begin{equation}
 H_{\rm
extended~hydro}\{\rho,{\bf v}\}= \int d^3x\left(\frac{1}{2}\rho v^2+
\epsilon(\rho) -\mu\rho + \frac{1}{2} K(\nabla\rho)^2\right)~,
\label{ExtendeClassicalHydrodynamic}
\end{equation}   
 The other possible terms are $\propto (\nabla\cdot{\bf v})^2$ and  
$\propto (\nabla\times{\bf v})^2$,
 which we do not discuss here.
While the Hamiltonian can be extended, the Poisson brackets for
hydrodynamics variables remain intact. It was also stressed by Landau that
hydrodynamic equations are less general than the commutation relations
for hydrodynamic operators.

\subsection{Classical superfluid hydrodynamics}
\label{HydrodynamicsOfSuperfluids}

Let us introduce the quantity
 \begin{equation}
 \bar\kappa\equiv\frac{\kappa}{2\pi}=2\sqrt{K\rho}~,
\label{kappa}
\end{equation}   
which has dimension of circulation of velocity.
Then the lassical superfluid hydrodynamics is obtained if one  considers
within the extended classical hydrodynamics the class of the potential 
velocity fields: 
\begin{equation}
{\bf v}=\bar\kappa\nabla \theta~.
\label{PotentialVelocity}
\end{equation}  
In this normalization the flow potential $\theta$ is dimensionless.  This
allows us to introduce instead of $\rho$ and $\theta$  the classical
complex field  where  the dimensionless  $\theta$ plays the role of the
phase: $\Psi=\sqrt{\rho}e^{\theta}$.   In terms of  $\Psi$ the extended
version of the hydrodynamic Lagrangian in Eq.(\ref{ClassicalLagrangian})
becomes:
\begin{equation}
L_{\rm
GP}(\Psi)=  \frac{i\bar\kappa}{2}\left(\Psi^*\partial_t
\Psi-\Psi\partial_t \Psi^*\right)
+\frac{\bar\kappa^2}{2}\nabla\Psi^*\nabla\Psi+
\epsilon(\rho)-\mu\rho~,~\rho=|\Psi|^2~.
\label{GPLagrangian}
\end{equation}

The Equation (\ref{GPLagrangian}) is the Lagrangian of  the famous
Gross-Pitaevskii (GP) theory generalized to the arbitrary function
$\epsilon(\rho)$. In the original Gross-Pitaevskii theory the
non-linear
term is quadratic,
$\epsilon(\rho)=(1/2)g\rho^2$, and the variation of $L_{\rm GP}(\Psi)$
leads to the nonlinear Schr\"odinger equation. Note that this nonlinear
Schr\"odinger equation (or the more general equation obtained using
the general form $\epsilon(\rho)$)  is the classical equation, since the
Planck constant $\hbar$ does not enter Eq.(\ref{GPLagrangian}). Instead one has the parameter
$\kappa$ (or $\bar\kappa=\kappa/2\pi$) which has the dimension
 of circulation of velocity $[\kappa]=[v][r]$. Circulation  $\oint d{\bf
r}\cdot{\bf v}$ is the adiabatic invariant in classical hydrodynamics,
and thus should be quantized in quantum theory.  Another invariant in
hydrodynamics is
$\int d^3x \left({\bf v}\cdot(\nabla\times {\bf v})\right)$. It is also quantized in quantum theory, see
Ref. \cite{Mineev}.

The superfluid hydrodynamics (SH) has three dimensional parameters ($c$, 
$\rho$ and $\kappa$), and thus the characteristic length, energy and
frequency scales are now determined:
\begin{equation}
 a_{\rm SH} =\frac{\bar\kappa}{c}~~,~~\omega_{\rm SH}=\rho\frac{\bar\kappa^3}{c^3} 
~~,~~E_{\rm SH}=\rho\frac{\bar\kappa^3}{c}~.
\label{ExtendedHydrodynamicScales}
\end{equation} 

The superfluid hydrodynamics is classical.  The
corresponding  hydrodynamic Hamiltonian is expressed in terms of the
classical velocity and mass density fields as in
Eq.(\ref{ExtendeClassicalHydrodynamic}):
 \begin{equation}
 H_{\rm
GP}\{\rho,{\bf v}\}=\int d^3x\left( \frac{1}{2}\rho v^2+
\epsilon(\rho)  -\mu\rho+ \frac{\bar\kappa^2}{8\rho}
(\nabla\rho)^2\right)~.
\label{HydrodynamicGP}
\end{equation}   
However, 
compared to the conventional classical hydrodynamics the classical
superfluid hydrodynamics described by Eq.(\ref{GPLagrangian}) has three
modifications: 

(i) The so-called quantum pressure term  $(\bar\kappa^2/8\rho)
(\nabla\rho)^2$ is added. In principle, this term can be of the classical
origin.  This term leads to the correction to the linear dispersion relation
for sound waves: $\omega(k)=ck\sqrt{1+\bar\kappa^2k^2/4c^2}$.

(ii)  The rotational degrees of freedom are involved in
this description. Since the phase $\theta$ is not single-valued, the
superfluid hydrodynamics (SH) contains vortices with quantized
circulation
$\oint d{\bf r}\cdot{\bf v}=n\kappa=2\pi n\bar\kappa$, where $n$ is
integer.  

(iii) Outside the vortex cores the velocity field is
potential, $\nabla\times{\bf v}=0$. 

The energy required to excite the vortex degrees of freedom is
the energy of the vortex loop $E_{\rm vr}\sim \rho\bar\kappa^2 r$ of
minimal size 
$r\sim a_{\rm SH}=\bar\kappa/c$ in Eq.(\ref{ExtendedHydrodynamicScales}).
Thus there is the gap for vortex excitations of order 
\begin{equation}
 \Delta\sim E_{\rm SH}=\frac{\rho\bar\kappa^3}{c}  ~.
\label{VortexGapEH}
\end{equation}  

The advantage of Lagrangian Eq.(\ref{GPLagrangian}) with the general 
function $\epsilon(\rho)$  compared to the
conventional  Ginzburg-Pitaevskii (GP) Lagrangian which describes
superfluid hydrodynamics in a dilute Bose condensate is as follows. 
In a dilute Bose gas almost all the atoms are in the Bose condensate, 
the depletion -- the difference between the total density of atoms and
the density of condensate is small and can be neglected in the main
approximation. As a result, the equation for the condensate practically
coincides with the hydrodynamics equations.  It should be mentioned that
the Ginzburg-Pitaevskii equation is not applicable to Bose condensate if
the depletion is not small, since there is no conservation law for the
condensate density.

For strongly interacting liquids the depletion is not small.  For example,
in superfluid $^4$He the condensate comprises only the small  fraction of
the total density. Nevertheless, even in this case, the
Eq.(\ref{GPLagrangian}) remains reasonable, since the function $\Psi$ is
normalized to the total density:
$|\Psi|^2=\rho$. This reflects the fact that the superfluid hydrodynamics 
describes not the dynamics of the condensate density, but the dynamics of
the whole superfluid liquid at $T=0$.

The Lagrangian in Eq.(\ref{GPLagrangian}) leads to correct hydrodynamic 
equations and to correct energy of quantized vortex lines both in the
dilute Bose gases and strongly interacting liquids. This implies that the
extended GP Lagrangian  gives the reasonable description of the classical
hydrodynamics of superfluids at $T=0$, which includes the hydrodynamics 
of superfluid component at $T=0$ and the classical dynamics of  vortices
with quantized circulation. The normal component made of quanta of sound
waves -- phonons -- is absent in this approach. It is included at the
stage of quantization to obtain the two fluid hydrodynamics at $T\neq
0$.  The drawback of this description is that as distinct from the GP 
equation for the dilute Bose gases, the general  Lagrangian in
Eq.(\ref{GPLagrangian}) gives only the model description of the vortex
core region; however, in many cases such model is sufficient since it
allows us to consider the core effects consistently without ambiguous
cut-off procedure. The further extension of the model with incorporation of the non-local
interaction can be found in Ref. \cite{Berloff}.

In conclusion, the model (\ref{HydrodynamicGP}) simulates superfluid
hydrodynamics not only in weakly interacting Bose gas, but also in real
quantum liquids, in which the Bose condensate is either absent or
is a small fraction of the total density. It is also important, that as 
distinct from the Bose gas,  liquids can be stable even in the absence of
environment, i.e. at zero external pressure. This is important for the
consideration of the problems of vacuum energy and the related problems
of cosmological constant \cite{Weinberg,Padmanabhan} using the ground
state of an isolated quantum liquid as the physical example of the
quantum vacuum in which the nullification of the vacuum energy in
equilibrium occurs without any  fine tuning \cite{Myths}.

\section{Quantum hydrodynamics}

\subsection{Landau quantum  hydrodynamics}

 Landau introduced quantum Hamiltonian  expressing the classical 
energy in Eq.(\ref{ClassicalHydrodynamicH}) it in terms of the
corresponding non-commuting quantum operators
$\hat{\bf v}$ and
$\hat\rho$:
\begin{equation}
\hat H_{\rm hydro}(\hat\rho,\hat{\bf v})= \int
d^3x\left(\frac{1}{2}\hat{\bf v}\hat\rho
\hat{\bf v}+
\epsilon(\hat\rho)-\mu\hat\rho\right)~.
\label{QuantumlHydrodynamicH}
\end{equation}  
The commutation relations for the components of velocity field
operator $\hat{\bf v}$, and between  $\hat{\bf v}$ and $\hat\rho$ are
\begin{eqnarray}
\left[\hat\rho({\bf r}_1),\hat\rho({\bf r}_2)\right]= 0~,
\label{CRDensityDensity}\\
\left[\hat{\bf v}({\bf r}_1),\hat\rho({\bf r}_2)\right]=\frac{\hbar}{i}
\nabla
\delta({\bf r}_1-{\bf r}_2)~,
\label{CRVelocityDensity}
\\
 \left[\hat v_{i}({\bf r}_1),\hat v_{j}({\bf
r}_2)\right]=\frac{\hbar}{i\hat\rho}e_{ijk}
 (\nabla\times \hat{\bf
v})_k \delta( {\bf r}_1-{\bf r}_2)~,
\label{CRVelocity}
\end{eqnarray}
have
been derived by Landau from the microscopics. They can also be obtained
from the Poisson brackets 
(\ref{PBDensityDensity})-(\ref{PoissonBracketsVelocity}) for the
classical variables.  

Quantum hydrodynamics is characterized  by  three dimensional quantities.
In addition to equilibrium values of $\rho$ and   $c$, the really 
fundamental Planck constant
$\hbar$ enters the quantum hydrodynamics through the commutation
relations (\ref{CRVelocityDensity}) and (\ref{CRVelocity}).

Using three dimensional 
quantities one can construct the characteristic `Planck' scales  
for  the energy $E_{\rm QH}$, mass $M_{\rm QH}$,  length $a_{\rm QH}$, 
 frequency
$\omega_{\rm QH}$ and energy density $\epsilon_{\rm QH}$:
 \begin{equation}
E_{\rm QH}^4=\frac{\hbar^3\rho}{c}~,~M_{\rm QH}^4=\frac{\hbar^3\rho}{c^3}~,
~a_{\rm QH}^4=\frac{\hbar}{\rho
c}~,~\omega_{\rm QH}=\left(\frac{c^5\rho}{\hbar}\right)^{1/4},
~\epsilon_{\rm QH}\sim
\epsilon(\rho)\sim \rho c^2~.
\label{HydrodynamicMass}
\end{equation} 
 
 \subsection{Rotational modes}

Landau suggested that the only low frequency modes of  quantum
hydrodynamics are quanta of sound waves -- phonons, while the rotational
modes (vortex degrees of freedom) are separated by the gap. One may
suggest that if the gap exists in quantum hydrodynamics, it is given by
the characteristic energy scale $E_{\rm QH}$ in
Eq.(\ref{HydrodynamicMass}). However, there are some arguments against
that.  Since the operators of vorticity $\nabla\times\hat{\bf v}$ and
density  $\hat\rho$ are commuted, the Hamiltonian which governs the
rotational degrees of freedom is 
\begin{equation}
\hat H_{\rm transverse}(\hat{\bf v}_\perp) =\frac{1}{2}\int
d^3x\rho\hat{\bf v}_\perp^2~,
\label{QuantumlHydrodynamicHTransverse}
\end{equation}  
where $\hat{\bf v}_\perp$ is the transverse (non-potential)  part of
velocity field. The above vortex contribution to quantum hydrodynamics
contains only two parameters: $\hbar$ and $\rho$. Using these two
quantities only  one cannot construct the quantity with the dimension of
the energy gap: the ``Planck'' energy scale $E_{\rm QH}$ of quantum
hydrodynamics in Eq,(\ref{HydrodynamicMass}) contains $c$ which is
irrelevant for transverse degrees of freedom.

This was probably the reason why Landau proposed different  estimate for
the rotational gap which did not contain $c$,   but contained the mass
$m$ of $^4$He atom: \cite{LandauQH} 
 \begin{equation}
\Delta_L=\frac{\hbar^2 \rho^{2/3}}{m^{5/3}}~,
\label{LandauGap}
\end{equation} 
The atomic mass  $m$ is the microscopic parameter,  which is beyond
the quantum hydrodynamics.  Incidentally or not,  but since in
superfluid $^4$He the atomic mass  $m$ and the quantum hydrodynamic
mass $M_{\rm QH}$ in Eq.(\ref{HydrodynamicMass}) are of the same order,
the Landau estimation in Eq.(\ref{LandauGap}) coincides with the
estimation for the energy of the elementary  vortex excitation in
superfluid $^4$He -- the smallest possible vortex ring -- in
Eq.(\ref{VortexGapEH}).

That quantum hydrodynamics alone cannot describe the superfluid  liquid
has been later emphasized by Feynman \cite{Feynman}. The main reason is
that the classical hydrodynamics lacks (has lost) the information on the
important microscopic properties of the underlying system, such as
quantum statistics of atoms. It is the Bose statistics of atoms which 
leeds to the gap in the spectrum of quantum vorticity \cite{Feynman}.
However, such gap is not present in  Fermi liquids (unless the Cooper
pairing occurs). Moreover, Fermi liquids are not described by classical
or quantum  hydrodynamics.  

All this demonstrates that in order to describe the real systems,  the
quantum hydrodynamics requires the extension. As the starting point for
quantization one can choose the extended classical hydrodynamics
discussed in Sec. \ref{HydrodynamicsOfSuperfluids}.  In this approach the
vortex degrees of freedom would have the gap already at the classical
level (see Eq.(\ref{VortexGapEH})). This will be
discussed later in Sec. \ref{ExtendedQH}.

\subsection{Quantization of phonon field}

If for some reasons the rotational degrees of freedom  are separated by
the gap, then the only low-energy degrees of freedom are represented by
the  vortex-free hydrodynamics and sound waves. In linear regime  (and in
the absence of the rotational degrees of freedom) the Landau quantum
hydrodynamics leads to quantization of sound waves. Quanta of sound waves
are phonons with linear spectrum $E_k=\hbar ck$. 

The nonlinear terms in  quantum hydrodynamic Hamiltonian  describe
interaction of phonon fields, and lead to modification of phonon spectrum
at large $k$. One may expect that the linear dispersion of phonon
spectrum (analog of Lorentz invariance) is violated at the Planck scale
$k_{\rm QH}=1/a_{\rm QH}$ in Eq.(\ref{HydrodynamicMass}), provided that no
microscopic physics intervenes earlier. 
In principle, the correction to
the spectrum  of phonons can be computed within the quantum
hydrodynamics, however the diverging Feynman diagrams  makes this
procedure  rather ambiguous. In the literature, people used the inverse inter-atomic 
distance $k_a\sim 1/a$ as the natural ultraviolet cut-off for diverging diagrams
(see e.g. \cite{Eckstein}). However,
such parameter characterizes the microscopic physics beyond the quantum hydrodynamics.
Whether it is possible to make regularization in such a way that the natural cut-off 
is determined by the quantum hydrodynamics itself,  i.e. by $k_{\rm
QH}=1/a_{\rm QH}$, is an open question.

 If  such regularization procedure exists, the first  guess would be that the spectrum is modified  by the next order term:
 \begin{equation}
\omega^2= c^2k^2(1+\gamma a_{\rm QH}^2k^2 + ...)~~,~~|\gamma|\sim 1~.
\label{ModPhononSpectrum4}
\end{equation} 
The quantum
hydrodynamic correction in the form of Eq.(\ref{ModPhononSpectrum4})  is
naturally obtained in the 1+1 case, when $\rho$ is the one-dimensional
mass density and one has $a_{\rm QH}^2=\hbar/(\rho c)$.  In case of
the 3+1 quantum hydrodynamics, where according to
Eq.(\ref{HydrodynamicMass})  one has $a_{\rm QH}^4=\hbar/(\rho c)$, the
correction in Eq.(\ref{ModPhononSpectrum4}) is non-analytic in $\hbar$,
and  the proper correction would be the higher order term which is linear
in
$\hbar$:
  \begin{equation}
\omega^2=  c^2k^2\left(1+\gamma a_{\rm QH}^4k^4
\ln\left(\frac{1}{a_{\rm QH}^4k^4}\right)  + ...\right)~~,~~|\gamma|\sim
1~.
\label{ModPhononSpectrum6}
\end{equation} 
The above guess is supported by the temperature  corrections to the
spectrum of phonons
\cite{Khalatnikov} after substitution $T\sim \hbar ck$. 
 
 The density of zero point energy of quantum  phonon field can be
estimated using the Planck energy cut-off $k_{\rm QH}=1/a_{\rm QH}$:
   \begin{equation}
\epsilon_{\rm zp}=\frac{1}{2}\int \frac{d^3k}{(2\pi)^3}  \hbar ck\sim
\hbar ck_{\rm QH}^4\sim \frac{E_{\rm QH}}{a_{\rm QH}^3} \sim \rho c^2~.
\label{ZPE}
\end{equation} 
This gives the correct estimation of at least the magnitude  of the
energy density of the liquid (the sign of $\epsilon(\rho) -\mu\rho$ is
negative if the external pressure is positive). Note that the result is
classical, i.e. it does not depend on $\hbar$. This is not very
surprising because   the energy density constructed from  $\hbar$, $c$
and $\rho$  does not contain $\hbar$ (see  Eq.(\ref{HydrodynamicMass})). 

From the modern point of view, the classical hydrodynamics  as well as
classical gravity, is the classical output of the quantum system in the
low-energy corner. The `initial classical'  energy density
$\epsilon(\rho)$ is not only  the starting point  for `quantum
hydrodynamics' but also is the final classical macroscopic result: it
contains all the quantum contributions to the energy density of the
liquid.  This also means that the contribution of zero point energy of
phonons to vacuum energy has already been included from the very
beginning and should not be counted again. Thus the phonon Hamiltonian
in the quadratic approximation must be written without zero-point 
energy of phonons:
\begin{equation}
\hat H  =  E_{\rm vac} + \sum_{\bf k}\hbar ck a^\dagger_{\bf k}a_{\bf
k}~~, ~~E_{\rm vac}=V\epsilon_{\rm vac}=V\left( 
\epsilon(\rho_0)-\mu\rho_0\right)~,
\label{Phonons}
\end{equation}  
where $a^\dagger_{\bf k}$ and $a_{\bf k}$  are operators of creation and
annihilation of phonons, and $V$ is the volume of liquid.

 \subsection{Quantum superfluid hydrodynamics}
 \label{ExtendedQH}

There are several ways of quantization of superfluid hydrodynamics:  (i)
One can perform the full Landau quantization of
Eq.(\ref{ExtendeClassicalHydrodynamic}), expressing the extended
hydrodynamic Hamiltonian in terms of the quantum fields $\hat\rho$ and
$\hat{\bf v}$. (ii) One can perform quantization starting with the
superfluid hydrodynamics with quantized vortices described by the
Lagrangian in Eq.(\ref{GPLagrangian}) by expressing it in terms of the 
non-commuting fields $\hat\Psi$ and $\hat\Psi^\dagger$.  Let us consider
the last case.
 The quantum counterpart of classical Lagrangian Eq.(\ref{GPLagrangian})  
is the Hamiltonian
 \begin{equation}
\hat H_{\rm
GP}(\hat\Psi)= \int d^3x\left(
\frac{\bar\kappa^2}{2}\nabla\hat\Psi^\dagger\nabla\hat\Psi+
\epsilon(\hat\rho)-\mu\hat\rho\right)~,~\hat\rho=\hat\Psi^\dagger\hat\Psi~,
\label{QuantumlHydrodynamicGP}
\end{equation}  
which is supplemented by commutation relations for quantum fields
  \begin{equation}
 \left[ \hat\Psi({\bf r}_1),\hat\Psi^\dagger({\bf r}_2)\right]=
\frac{\hbar}{\bar\kappa}\delta({\bf r}_1-{\bf r}_2)~.
\label{Commutations}
 \end{equation} 
If one identifies the parameter ${\hbar}/{\bar\kappa}$ with the  mass  of
an atom of the liquid, one obtains that the extended quantum
hydrodynamics is nothing but the microscopic quantum mechanics of a
system of identical bosonic atoms with mass $m={\hbar}/{\bar\kappa}$ and
with a special type of interaction term $\epsilon(\hat\rho)$ which only
depends on density.

The superfluid quantum hydrodynamics (SQH) contains four  parameters
$\hbar$,
$m$,  speed of sound $c$, and equilibrium density $\rho$. One can
introduce the dimensionless mass 
 parameter $m_{\rm SQH}$:
 \begin{equation}
m_{\rm SQH}=\frac{m}{M_{\rm QH}}~.
\label{Dimensionlessparameter}
\end{equation}  
One may suggest that this dimensionless parameter   characterizes
microscopically different systems, which have the common macroscopic
(low-energy, hydrodynamic) properties. In dilute  Bose gases one 
has $m_{\rm SQH}\ll 1$, while in superfluid liquid $^4$He and in
superfluid liquid
$^3$He this parameter is of order unity, $m_{\rm SQH}\sim 1$.

\begin{figure}
 \centerline{\includegraphics[width=0.9\linewidth]{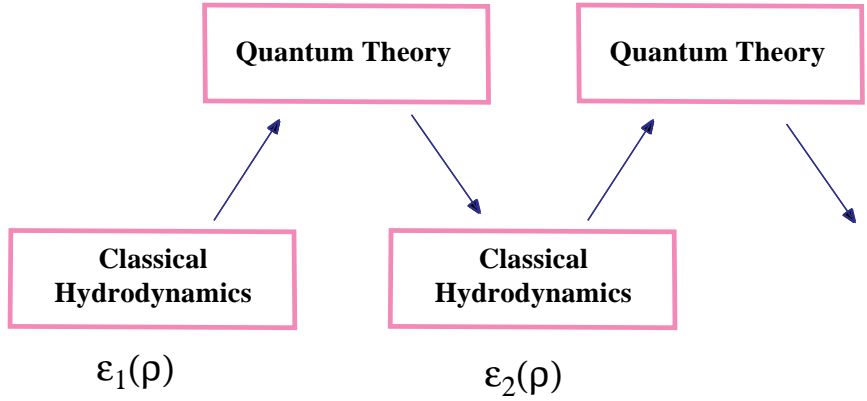}}
  \caption{From classical to quantum hydrodynamics (quantization) and back (to low-energy limit of quantum system).}
  \label{UpDownFig}
\end{figure}

However, if one compares the quantum hydrodynamic Hamiltonian
(\ref{QuantumlHydrodynamicGP})  with the Hamiltonian of exact microscopic
theory 
\begin{eqnarray}
H_{\rm micro}= \int d^3x \hat\Psi^\dagger({\bf
x})\left(-\frac{\bar\kappa^2}{2}\nabla^2  -\mu
\right) \hat\Psi({\bf x}) 
\nonumber
\\
+{1\over 2}\int d^3x\int d^3y ~\hat\Psi^\dagger({\bf x})
 \hat\Psi^\dagger({\bf y})U({\bf x}-{\bf
y}) \hat\Psi({\bf y}) \hat\Psi({\bf x}),
\label{TheoryOfEverything}
\end{eqnarray}
one finds that the difference in the interaction term is enormous. 
In other words, the prescribed down-up route 
from classical  to quantum theory (see Fig. \ref{UpDownFig}) does not
lead in general to the true microscopic theory. 

And this is not the only
drawback of quantum hydrodynamics. One may suggest that inspite of
disagreement with exact microscopic theory, the `microscopic'
Hamiltonian in Eq.(\ref{QuantumlHydrodynamicGP}) may serve as a relevant
microscopic model. In principle, starting with this Hamiltonian, one may
obtain in the long-wave limit (i.e. in the up-down route in Fig. 
\ref{UpDownFig}) the classical hydrodynamic Hamiltonian for superfluid
liquid state. However, the emerging function
$\epsilon$  will essentially deviate from
$\epsilon$ in the original classical  hydrodynamics, i.e.
$\epsilon_2(\rho)\neq \epsilon_1(\rho)$.  Moreover this function
$\epsilon(\rho)$ cannot be expressed in terms of the renormalized
coupling $g$. That is why this procedure -- down-up (quantization),
up-down (emergence of effective theory in the low-energy corner of
quantum theory), down-up, etc. in  Fig. \ref{UpDownFig} --  in
general does not converge.


 \subsection{Weak coupling limit}
 \label{WeakCoupling}

The only case in which the exact  theory and 
extended quantum hydrodynamics fit each other is when the energy density 
is quadratic function of $\rho$: 
 \begin{equation}
  \epsilon(\hat\rho)=\frac{g}{2}(\hat\Psi^\dagger\hat\Psi)^2 ~,
\label{QuadraticInteraction}
\end{equation} 
and the parameter $m_{\rm SQH}$ in Eq.(\ref{Dimensionlessparameter})  is
small: $m_{\rm SQH}\ll 1$. This corresponds to small coupling $g$ 
  \begin{equation}
\frac{g}{g_0} =m_{\rm SQH}^{8/3}\ll 1~~,~~
 g_0=\frac{\hbar^2a}{m^3}=\frac{\hbar^2}{\rho^{1/3}m^{8/3}} ~.
\label{SmallCoupling}
\end{equation} 
  where $a=(\rho/m)^{-1/3}$ is the interatomic distance.
In liquid
$^4$He one has $m\sim M_{\rm QH}$, and thus  
quantization of the hydrodynamics does not make sense.

The limit of small coupling $g$ corresponds to the model of  weakly
interacting Bose gas, which has been solved by Bogoliubov. When one
starts with the microscopic theory 
 \begin{equation}
\hat H_{\rm
GP}(\hat\Psi)= \int
d^3x\left(\frac{\hbar^2}{2m^2}\nabla\hat\Psi^\dagger\nabla\hat\Psi+
\frac{g}{2}\hat\rho^2-\mu\hat\rho\right)~ 
\label{MicroscopicGP}
\end{equation}  
with small $g$,  one obtains  in the long-wave-length limit (i.e. on the
up-down route in   Fig. \ref{UpDownFig}) the classical
hydrodynamics, where in the main
approximation  the function
$\epsilon(\rho)=(g/2)\rho^2$ coincides with that in microscopic theory.
In the next approximation the function $\epsilon(\rho)$ is modified by
the ``quantum'' correction, as follows from the Bogoliubov theory:
 \begin{equation}
\epsilon(\rho)=\frac{g}{2}\rho^2\left(1 +
\frac{16}{15\pi^2}\left(\frac{g}{g_0}\right)^{3/2}  \right) ~~,~~g\ll g_0 
  ~.
\label{CorrectionEpsilon}
\end{equation} 
This means that after the first iteration (down-up and up-down in Fig.
\ref{UpDownFig}) the coupling constant is renormalized:
 \begin{equation}
\tilde g= g\left(1 +
\frac{16}{15\pi^2}\left(\frac{g}{g_0}\right)^{3/2}  \right) ~~,~~g\ll g_0 
  ~.
\label{FirstItreation}
\end{equation}

There is a temptation to consider the correction to
$\epsilon(\rho)=(g/2)\rho^2$ in Eq.(\ref{CorrectionEpsilon}) as the back
reaction of the quantum vacuum to quantum fields of phonons
\cite{Fischer}. At first glance,  one may identify this correction with
the properly regularized zero point energy of phonon field:
\begin{eqnarray}
\epsilon_{\rm zp}=\frac{1}{2}\int
\frac{d^3k}{(2\pi)^3}  \hbar ck\left(\sqrt{1+\frac{\bar\kappa^2k^2}{4c^2}} - \frac{\bar\kappa k}{2c}- \frac{c}{\bar\kappa k}+
\frac{c^3}{\bar\kappa^3k^3} \right) 
\label{counter-terms}
\\
=\frac{8}{15\pi^2}\frac{\hbar c^5}{\bar\kappa^4}  
=\frac{8}{15\pi^2}g\rho^2
\left(\frac{g}{g_0}\right)^{3/2}   
  ~.
\label{CorrectionAsZP}
\end{eqnarray} 
 However, such interpretation is only valid at small  $g$ when the
microscopic Bogoliubov theory provides three counter-terms  in
Eq.(\ref{counter-terms}).

Moreover, this correction contains the Planck constant  $\hbar$ in the
denominator  (since
$\bar\kappa=\hbar/m$). This means that the  weakly interacting Bose gas
(the system with small $g$) actually corresponds to the ultra-quantum
limit, in which the contribution of zero point motion of phonon field is
small compared to the main quantum contribution
$(1/2)g\rho^2$ to the vacuum energy.

 \subsubsection{Energy Scales}

Because of the dimensionless quantity $g/g_0$ (or $m/M_{\rm QH}$), the
 extended quantum hydrodynamics in Eq.(\ref{MicroscopicGP}) contains
different physically interesting scales for each dimensional quantity.
In addition to the hydrodynamic energy  scale
$E_{\rm QH}=M_{\rm QH}c^2$ in Eq.(\ref{HydrodynamicMass}),  there is the
scale
$E_{\rm L}=mc^2\ll E_{\rm QH}$,  where  the Lorentz violation occurs in
a  dilute Bose condensate.
 
Another energy scale was introduced by
Landau \cite{LandauQH} 
in Eq.(\ref{LandauGap}). This corresponds to the energy of the smallest
vortex ring in superfluid
$^4$He: $E_{\rm vr}\sim \rho \bar\kappa^2 a \sim \Delta_L$, where
$a$  is the interatomic
spacing ($\rho\sim m/a^3$) which determines the smallest possible
radius of the vortex ring in superfluid $^4$He. 
 
 There are several  important length scales in dilute Bose gases  in
addition to the
 hydrodynamic length $a_H$: 
 interatomic spacing $a$ and the coherence length  $\xi=\hbar/mc$:
 \begin{equation}
 \frac{a_H}{a}= \left(\frac{g_0}{g}\right)^{1/8}~, ~ \frac{\xi}{a}= 
\left(\frac{g_0}{g}\right)^{1/2}.
\label{lengthes}
\end{equation}

 \section{Back reaction of quantum vacuum}
\subsection{Depletion of condensate}

The term `depletion' means that because of interaction (i.e.
for $g\neq 0$) the non-vanishing number of atoms is not in the
Bose-condensate. For a weakly interaction dilute Bose gas, the atoms in
the condensate are prevailing. The relative values of the condensate and
non-condensate mass densities are
 \begin{equation}
~~\frac{\rho_{\rm cond}}{\rho} =1
-\frac{1}{3\pi^2}\left(\frac{g}{g_0}\right)^{3/2} ~~,~~
\frac{\rho_{\rm non-cond}}{\rho}
=\frac{1}{3\pi^2}\left(\frac{g}{g_0}\right)^{3/2}  ~.
\label{CondensateDepletion}
\end{equation} 
In strongly interacting $^4$He liquid the fraction of the non-condensate
atoms is prevailing, with $\rho_{\rm cond}/\rho< 0.1$. Nevetheless, at
$T=0$ the whole liquid is in the coherent superfluid state  -- the
quantum vacuum -- with the superfluid component density $\rho_{\rm
s}=\rho$. The same occurs in two dimensional systems, where the
condensate is completely depleted even in the
absence of interaction, $\rho_{\rm cond}=0$.

The depletion of the Bose-condensate is not in the framework of Landau
quantum hydrodynamics. It is in the framework of the Tisza description
of superfluids and is fully microscopic
phenomenon, which is beyond the low-energy hydrodynamics. 
Let us stress again that the Landau description in
terms of vacuum and matter (quasiparticles) is applicable 
for superfluids in the low-energy regime.  In this regime the
hydrodynamics with its Euler and continuity equations has no information
on the separation of the liquid into the Bose condensate and atoms above
the condensate caused by interaction, since at
$T=0$ both these fractions participate in a single coherent flow of the
quantum vacuum. The Tisza picture of condensed and
non-condensed fractions requires the microscopic description of
the particle and energy exchange between the two fractions; this is
the high-energy phenomenon which is certainly beyond the responsibility
of hydrodynamics.

In general, the depletion of the Bose-condensate is also beyond the
quantum superfluid hydrodynamics, except for the limit $g\ll g_0$, where
the superfluid quantum hydrodynamics coincides with the microscopic
Bogoliubov model, and the depletion can be studied using perturbation
theory.  This is the reason why the calcuations of the depletion using 
the quantum fluctuations of phonon field  (see e.g. 
Ref.\cite{Fagnocchi}) or other back reaction effects (see e.g. Ref.
\cite{Fischer}) cannot be considered as generic.

However, there are some problems which are within the responsibility of 
Landau quantum hydrodynamics. One of them is the depletion of the mass
density caused by phonons. This is the back reaction of quanta of sound
waves onto the `classical' quantum vacuum (let us stress again that in the
low-energy limit the superfluid quantum vacuum behaves as classical
liquid).

\subsection{Back reaction of vacuum density to quantum matter}
\label{BackReactionDensity}

At non-zero temperature the liquid consists of the vacuum (the ground
state) with density $\rho$ and excitations (quanta of sound waves --
phonons)  in Eq.(\ref{Phonons}). Lets us find how thermal phonons modify
the mass density $\rho$ of the quantum vacuum. This is the back reaction
of the vacuum to the quanta of sound waves. We assume that temperature is
small,
$T\ll E_{\rm QH}$, so that only low-frequency  phonons with linear
spectrum $\omega=ck$ contribute to the thermal energy, and consider 
fixed external pressure. The correction can be obtained by  minimization
of the free energy density of the liquid
$F=\epsilon-TS$ over
$\rho$. The free energy is the sum of the  energy of ground state 
(quantum vacuum)  and the free energy of the phonon gas (matter).  For
the phonons with linear dispersion relation the free energy density is
the radiation pressure with minus sign: 
\begin{equation}
F_{\rm mat}= - P_{\rm mat}=-(1/3)\epsilon_{\rm
mat}~~,~~ \epsilon_{\rm mat}=\frac{\pi^2}{30 \hbar^3c^3}T^4~,
\label{RadiationEnergy}
\end{equation} 
where $\epsilon_{\rm mat}$ is the energy density of the gas of thermal
phonons (radiation energy).

Since the vacuum does not contribute to the entropy of the system, the
total free energy density of a liquid is
\begin{equation}
F (T,\rho)=\epsilon(\rho) -\mu \rho- \frac{1}{3}\epsilon_{\rm
mat}(\rho)~.
\label{FreeEnergy1}
\end{equation} 
Let $\rho_0$ be the equilibrium density at $T=0$ and $\mu=\mu_0$, then
considering $\epsilon_{\rm mat}$ as perturbation one obtains the
following expansion in terms of $\delta\rho=\rho-\rho_0$ and
$\delta\mu=\mu-\mu_0$:
\begin{equation}
F (T,\rho)=F (T,\rho_0)+\frac{1}{2}\frac {\partial^2  \epsilon_{\rm
vac}}{\partial\rho^2}(\delta \rho)^2 - \frac{1}{3} \frac {\partial
\epsilon_{\rm mat}}{\partial\rho}\delta \rho-\delta\mu\delta\rho~.
\label{FreeEnergy}
\end{equation} 
For
phonon gas the dependence of the radiation energy  on $\rho$  in
Eq.(\ref{RadiationEnergy}) only comes  from the speed of sound, 
\begin{equation}
\frac{\partial \epsilon_{\rm mat}}{\partial\rho}=-3\frac{\epsilon_{\rm
mat}}{c}\frac{\partial   c}{\partial   \rho}=-3u\frac{\epsilon_{\rm
mat}}{\rho}~.
\label{cofrho}
\end{equation} 
Here we introduced the function $u$
\begin{equation}
  u=\frac{\partial \ln c}{\partial \ln \rho}~,
\label{Gruneisen}
\end{equation} 
 which is the fluid-state analogue
of Gr\"uneisen parameter,  see e.g.
\cite{Khalatnikov}.

Then we must take into account that the
chemical potential $\mu$ must be changed to support the fixed external
pressure. The total change of the pressure of the liquid,  which is the
sum of the vacuum pressure of the liquid and the radiation pressure of
phonons, must be zero, $\delta P_{\rm vac}+ P_{\rm mat}=0$. This gives
\begin{equation}
 \delta P_{\rm vac}=- P_{\rm mat}=
-\frac{1}{3}\epsilon_{\rm mat} ~,
\label{PressureBalance}
\end{equation} 
As a result the change in the chemical potential is
\begin{equation}
\delta\mu=\frac{\delta P_{\rm vac}}{\rho_0}=
-\frac{1}{3}\frac{\epsilon_{\rm mat}}{\rho_0}  ~.
\label{deltamu}
\end{equation}

Introducing Eqs. (\ref{cofrho}) and (\ref{deltamu}) into free energy
(\ref{FreeEnergy}) and minimizing over $\delta\rho$  one obtains the
response of the density of the liquid to the phonon gas:
\begin{equation}
\frac{\delta\rho}{\rho}= -\frac{\epsilon_{\rm
mat}}{\rho c^2}\left(\frac{1}{3} + u\right)~.
\label{DensityDepletion}
\end{equation}

The result in Eq.(\ref{DensityDepletion}) can be also obtained from the
analysis of classical hydrodynamic equations made by Stone in Refs.
\cite{Stone2000,Stone2002}.  The second term on the rhs of 
Eq.(\ref{DensityDepletion})  comes from the second order correction to
the density of the liquid induced by the sound wave. This is the
Eq.(4.13) of Ref.\cite{Stone2000} integrated over thermal quanta
of sound waves -- phonons.  The first term in
the rhs of Eq.(\ref{DensityDepletion}), which is due to the change in the
vacuum pressure, can be also obtained using Stone's formalism.

Note that the depletion of liquid density   $\delta\rho\propto T^4$,
 while the temperature correction to the depletion of the condensate is
$\propto T^2$ (see e.g. \cite{Fagnocchi}) . The reason for such
difference is that the density $\rho$ is conserved quantity, while the
condensate density is not  because of the Josephson coupling between the
condensate and non-condensate atoms. In conclusion, the depletion of the
mass density is universal and is completely determined by hydrodynamics,
while the depletion of the condensate is beyond the quantum
hydrodynamics and strongly depends on the microscopic physics.

\subsection{Response of dark (vacuum) energy to matter}

Let us consider the back reaction of vacuum energy to thermal
phonons. According to Eq.(\ref{EquilibriumLiquid}) the analog of the
vacuum energy density in liquids is 
 $\epsilon_{\rm vac} =\tilde\epsilon=\epsilon(\rho) -\mu\rho$. It obeys
the correct equation of state for quantum vacuum  
\begin{equation}
P_{\rm vac}= \mu\rho-\epsilon(\rho)= -\epsilon_{\rm vac} ~.
\label{VacEqState}
\end{equation}  
The correction to the `vacuum energy' density due to thermal phonons is
\begin{equation}
\delta \epsilon_{\rm vac} =\delta(\epsilon(\rho) -\mu\rho)=
\left(\frac{d\epsilon}{d\rho} -
 \mu_0\right)\delta\rho- \rho_0\delta\mu
=- \rho_0\delta\mu
= \frac{1}{3}\epsilon_{\rm mat}~,
\label{SpecialRel}
\end{equation}  
where we used Eq.(\ref{deltamu}) for $\delta\mu$. 

Let us consider an equilibrium liquid in the absence of environment,
i.e. when the external pressure is zero. Then in the absence of phonons 
the vacuum energy and pressure are zero according to
Eq.(\ref{Nullification}), $P_{\rm vac}= 
-\epsilon_{\rm vac}=0$. At  $T\neq 0$, thermal phonons produce radiation
pressure which must be compensated by the pressure of the vacuum. As a
result the vacuum energy density becomes non-zero:
\begin{equation}
 \epsilon_{\rm vac} = \frac{1}{3}\epsilon_{\rm mat}~.
\label{SpecialRel2}
\end{equation}  
This is the back reaction of the vacuum to relativistic matter. The  same
relation between the dark energy and hot matter is
applicable for such Universes in which gravity is absent, i.e. in which
the Newton constant $G=0$  (see Refs. 
\cite{SpecialRel,Book}).  Note that in liquids, where the
effective gravity obeys hydrodynamic equations rather than Einstein
general relativity, the vacuum energy is naturally of the order of matter
density. For Universes with gravity, situation is more compicated, since
the vacuum energy responds also to gravitating matter, curvature,
expansion and other perturbations of the vacuum state. However, the main
result is that the vacuum energy is naturally determined by macroscopic
quantities, rather than by huge microscopic Planck energy scale
\cite{Myths}.

 \section{Lessons for quantum  gravity}

 \subsection{From quantum  gravity to quantum hydrodynamics}

The results in Eqs.(\ref{DensityDepletion}) and (\ref{SpecialRel}) for
the back reaction of the vacuum are expressed completetely in terms  of
quantum hydrodynamics, i.e.  in terms of the function $\epsilon(\rho)$ 
and Planck constant $\hbar$. These results are generic and do not
depend on the microscopic physics, so that the extension to quantum
superfluid hydrodynamcs (with its extra parameter
$\bar\kappa=\hbar/m$) is not required. The only role of the microscopic
physics is to supply us with the macroscopic function
$\epsilon(\rho)$. 
 
In general relativity, there are also examples of the universal
behavior of the back reaction, such as 
universal temperature corrections  to Einstein equations and to Newton
constant
$G$. The
temperature correction to the free energy of gravitational field induced
by $N_F$ massless fermionic quantum fields and $N_s$ scalar quantum
fields is \cite{Gusev-Zelnikov} 
\begin{equation}
F=  \frac{N_F-2N_s}{288\hbar} \int d^3x \sqrt{-g} T^2 [{\cal R}+6 w^2]
~.
\label{GR}
\end{equation} 
Here ${\cal R}$ is the Ricci curvature of gravitational field  and  
$w^2=w^\mu w_\mu$, where $w_\mu={1\over 2}\partial_\mu\ln g_{00}$  is
4-acceleration. This result also does not depend on the microscopic
Planck physics. It is expressed in terms of the Planck constant $\hbar$ 
and integral numbers  -- numbers of species $N_F$ and $N_s$ (actually one
should also add contribution of the vector fields and gravitons). Thus
the  only role of the microscopic Planck physics  is to supply us with
the definite number of fermionic and bosonic quantum fields in the low
energy corner. 

Moreover, the temperature correction to the
gravitational action in Eq.(\ref{GR}) is applicable not only to
 general relativity but also to the effective gravity
emerging in quantum liquids  \cite{Volovik-Zelnikov}. In
quantum liquids, the dominating  contribution to the `gravitational
action' is provided by hydrodynamics, while the subdominant corrections
are within responsibility of the QFT in curved space.  For superfluid
$^4$He and for Bose condensate of single atomic species the microscopic
physics  gives us $N_F=0$ and
$N_s=1$. Expressing ${\cal R}$,
$g$ and $w$ in Eq.(\ref{GR}) in terms of the effective  metric
$g_{\mu\nu}$ experienced by phonon field in
Eq.(\ref{CovarianAcousticMetric}),
one obtains the correct 
subdominant contribution to the hydrodynamic free energy of the liquid
$^4$He or Bose gas. In case of effective gravity in superfluid $^3$He-A
with gapless fermions,  the microscopic physics  gives us  $N_F=2$ and
$N_s=0$, and using Eq.(\ref{GR}) one obtains the correct subdominant
contribution to the gradient energy. These are examples when general
relativity helps us to solve some problems in superfluids.

 Another example is provided by the universal quantum correction to
Newton law (see e.g. \cite{Kirilin}). It has exact analog in quantum
hydrodynamics and gives rise to the universal quantum correction to the
classical hydrodynamic action caused by effective QFT in effective
curved space of acoustic metric (see Refs.
\cite{Ilinski}). As an illustration let us write one of the typical terms
generated by the quantum hydrodynamics -- the contribution to the
quantum pressure caused by quantum fluctuations of phonon
field in effective curved acoustic space obtained by
Seeley-De Witt expansion \cite{Ilinski}:
\begin{equation}
P_{\rm quantum}\sim
\hbar c~(\nabla^2\ln\rho)^2 \ln \frac{E_{\rm QH}}{E_{\rm IR}} 
~.
\label{HydrodQuantumPressure}
\end{equation} 
This leads to the quantum correction to the spectrum of phonons in
Eq.(\ref{ModPhononSpectrum6}) which is proportional to $\hbar$. The
infra-red (IR) logarithmic divergence of the quantum hydrodynamic
corrections suggests that they may describe the creation of phonons
(matter) by the time dependent flow (gravitational field) in exact
analogy with particle production in gravitational field (see e.g. Ref.
\cite{Dobado}). In a similar way, in superfluid
$^3$He-A the logarithmically divergent action for the effective
electromagnetic field leads to the Schwinger-type production of
fermionic quasiparticles by the time-dependent order parameter
\cite{Exotic}. 

The
quantum pressure produced by quantum fluctuations of phonon
field in effective curved acoustic space-time is more pronounced in the
1+1 quantum hydrodynamics. The effect is related to the
gravitational trace anomaly in the 1+1 space-time
\cite{Balbinot}, and leads to the quantum correction to the phonon
spectrum which is also  
proportional to $\hbar$:  the factor in Eq.(\ref{ModPhononSpectrum4}) is
$\gamma a_{\rm QH}^2=-\hbar/(48\pi\rho c)$. 

The Hawking radiation also does not distinguish between
gravity obeying the general relativity and  effective gravity in liquids
obeying hydrodynamic equations
\cite{Unruh,Horizons}.  In both cases, Hawking radiation from an
astronomical or acoustic black hole is described as the
process of  semi-classical tunneling between (quasi)particle trajectories inside and outside the horizon
\cite{TunnelingVolovik,Parikh}.

 \subsection{From quantum hydrodynamics to quantum  gravity}

We considered some cases when the quantum hydrodynamics 
and quantum gravity allow us to obtain
the true corrections to hydrodynamics or/and to general relativity. 
There are some other examples of such kind, when the quantum
hydrodynamics and quantum gravity work. However, it is not the
general case. Quantum hydrodynamics and quantum gravity reproduce  only
those (mostly subdominant) terms in the action or in free energy which
do not contain dimensional parameters, such as Eqs. (\ref{GR}) and
(\ref{HydrodQuantumPressure}). In general, the down-up route from
classical  to quantum hydrodynamics (see Fig.
\ref{UpDownFig}) leads to the theory which does not  coincide with the
true microscopic theory.  This reflects the main property of the emegent
physics: there are only very few  up-down ways, i.e.  from the high
energy microscopic theory to  the low-energy macroscopic hydrodynamic
theory. The way depends on the universality class and is unique for
given universality class. But there are infinitely many down-up routes
from macroscopics to microscopics. This is the main message for those
who would like to quantize gravity and hydrodynamics. 

One can quantize sound waves in hydrodynamics to
obtain quanta of sound waves --  phonons \cite{LandauQH}. Similarly one
can quantize gravitational waves in general relativity to obtain
gravitons \cite{Bronstein}. But one should not use the low-energy
quantization for calculation of the radiative corrections which contain
Feynman diagrams with integration over high momenta. In particular, the
effective field theory  is not appropriate for calculations
of the vacuum energy in terms of the zero-point energy of quantum
fields. Such attempts lead to the cosmological
constant problem in gravity
\cite{Weinberg,Padmanabhan}, and to the similar paradox for the
vacuum energy in quantum hydrodynamics: in both cases the
vacuum energy estimated using the effective theory is by many
orders of magnitude too big.  We know how this paradox is
solved in quantum liquids \cite{Myths}, and we may expect that the same
general arguments based on the thermodynamic stability of the
ground state of the quantum liquid are applicable to the  
vacuum of relativistic quantum fields.

Another hint from hydrodynamics is that the underlying microscopic 
theory of quantum gravity must contain additional parameter to $\hbar$, 
$c$ and $G$. Then one has the dimensionless parameter, which
distinguishes between different microscopic theories with the same
macroscopic phenomenology. Example of such parameter in quantum
hydrodynamics  is  $m_{\rm SQH}$ in Eq.(\ref{Dimensionlessparameter}).
It appears that properly formulated quantum hydrodynamics makes sense
only in the limit when this parameter is small, i.e. for the case of
dilute Bose gases.  The necessity of the small parameter for the
emergent general relativity and/or gauge fields is emphasized by
Bjorken: `the emergence can only work if there is an extremely small
expansion parameter in the game' \cite{Bjorken}. The role of the small
parameter could be played by the ratio $E_{\rm Planck}/E_{\rm Lorentz}$
between the Planck energy scale and the energy scale above which the
Lorentz invariance is violated (see e.g. discussion in  Ref.
\cite{KlinkhamerVolovik}). 

As follows from the experience with different quantum condensed
matter systems,  the metric field $g_{\mu\nu}$ may naturally emerge in the low-energy corner of quantum vacuum. It is important that in some systems gravity emerges as effective geometry, rather than the spin-2 field. Even in such caricature gravity as the effective gravity for sound waves propagating in inhomogeneous moving liquids, the acoustic metric  $g_{\mu\nu}$ in Eq.(\ref{CovarianAcousticMetric}) is the emerging geometrical object, which has nothing to do with the spin-2 field.
Depending on the hierarchy of parameters of the underlying microscopic system (quantum vacuum), the geometry (metric field) may obey the nonlinear hydrodynamic equations, or  the nonlinear equations  of general relativity, or  Gross-Pitaevskii equations, etc. 

In some vacua gravity emerges
together  with all the ingredients of Standard Model: relativistic
chiral fermions and quantum gauge fields. This is the general
low-energy property  of vacua with the so called Fermi point in
momentum space \cite{FrogNielBook,Book,Horava}, which demonstrates that
gravity is the natural part of physics, and it should not be separated
from the other fermionic and bosonic classical and quantum fields. The
separation only occurs at low energy, because of the difference between
the running couplings for gauge fields and gravity. This means that if
gravity is the energent phenomenon, it should naturally emerge together
and simultaneously with the other physical fields and physical laws. This
is the main requirement for the future theory of quantum gravity.

\section*{Acknowledgments}
This work was
supported in part by the Russian Ministry of
Education and Science, through the Leading Scientific School
grant $\#$2338.2003.2, by
ESF COSLAB Programme and by the Russian Foundation for
Basic Research (grant 06-02-16002-a).


\vfill


\begin{thebibliography}{00}

\bibitem{Dreyer} O. Dreyer,  Emergent General Relativity,  gr-qc/0604075.


\bibitem{Book} G. E. Volovik, {\it The Universe in a Helium
Droplet} (Clarendon Press,  Oxford, 2003).


\bibitem{Babichev} E. Babichev, V. Mukhanov and A. Vikman, 
Escaping from the black hole? {\it  JHEP} {\bf 0609}, 061 (2006); 
hep-th/0604075.


\bibitem{Fagnocchi} R. Balbinot, S. Fagnocchi and A. Fabbri, The depletion
in Bose Einstein condensates using Quantum Field Theory in curved space,
cond-mat/0610367.


\bibitem{Cano} C. Barcelo, A. Cano, L.J. Garay, G. Jannes, 
Stability analysis of sonic horizons in Bose-Einstein condensates, {\it
Phys. Rev.} {\bf D74}, 024008, (2006).


\bibitem{Cherubini}  C. Cherubini, F. Federici, S. Succi, M. P. Tosi,
Excised acoustic black holes: the scattering problem in the time domain,
{\it Phys.Rev.} {\bf D72} 084016 (2005).


\bibitem{Klippert}  V.A. De Lorenci, R. Klippert, and D.H. Teodoro,
Birefringence in nonlinear anisotropic dielectric media,
{\it Phys. Rev.} {\bf D 70}, 124035 (2004). 


\bibitem{Rosquist}  K. Rosquist, A moving medium simulation of
Schwarzschild black hole optics, {\it  Gen. Rel. Grav.} {\bf 36},
1977--1982  (2004); gr-qc/0309104.
 
\bibitem{Horizons} G.E. Volovik,  Horizons and ergoregions in superfluids, 
  {\it J. Low Temp. Phys.} {\bf 145}, 337--356 (2006); gr-qc/0603093.


\bibitem{Hu} B. L. Hu, New View on Quantum Gravity and the Origin of the Universe, gr-qc/0611058.

\bibitem{LandauQH} L.D. Landau, Theory of superfluidity of helium-II, 
 {\it J.
Phys. USSR}, {\bf 5}, 71 (1941).


\bibitem{Bronstein} M. Bronstein, Quantentheorie schwacher
Gravitationsfelder, {\it Phys. Ztschr. der Sowjetuion}, {\bf 9}, 140--157
(1936); see also G. Gorelik,  Matvei Bronstein and quantum gravity: 
70th anniversary of the unsolved problem, {\it Physics-Uspekhi} {\bf 48}, 1039--1053 (2005).


\bibitem{Tisza} L. Tisza, {\it J. Phys. Radium} {\bf 1}, 164 (1940); {\it
ibid.}  {\bf 1}, 350 (1940).


\bibitem{Feynman} R.P. Feynman,  {\it Statistical Mechanics} (Benjamin,
Massachusetts, 1972).


\bibitem{DzyaloshinskiiVolovick}  I.E. Dzyaloshinskii  and
G.E. Volovick, Poisson brackets in condensed matter,  {\it Ann. Phys.}
{\bf 125}, 67--97 (1980);  


\bibitem{PoissonVorticity} G.E. Volovik and V.S. Dotsenko,  Poisson
brackets and continual dynamics of the vortex lattice in rotating HeII,
 {\it JETP Lett.} {\bf 29}, 576 - 579 (1979); G.E. Volovik,   Poisson
brackets scheme for vortex dynamics in superfluids and superconductors
and effect of band structure of crystal,  {\it JETP Lett.} {\bf 64 }, 
845--852  (1996);  cond-mat/9610157.

\bibitem{Rasetti} M. Rasetti  and  T. Regge,  {\it   Physica}   {\bf
A80}, 217 (1975).

\bibitem{Stone2000} M. Stone, Acoustic energy and momentum in a moving
medium, {\it Phys. Rev.} {\bf E62}, 1341--1350 (2000).


\bibitem{Stone2002} M. Stone, Phonons and forces: momentum
versus pseudomomentum in moving fluids, in: {\it Artificial Black
Holes}, eds. M. Novello, M. Visser and G. Volovik (World
Scientific, 2002), pp. 335--364.


\bibitem{Unruh} W.G. Unruh, Experimental black-hole
evaporation?  {\it Phys. Rev. Lett.} {\bf 46}, 1351--1354 (1981);
Sonic analogue of black holes and the effects of high
frequencies on black hole evaporation, {\it Phys. Rev.} {\bf
D51}, 2827--2838 (1995).


\bibitem{BLV} C. Barcelo, S. Liberati and M. Visser,  Analogue Gravity,
{\it  Living Rev. Rel.} {\bf 8}, 12 (2005); gr-qc/0505065. 


\bibitem{FrogNielBook} C.D. Froggatt   and  H.B. Nielsen,
{\it Origin of Symmetry} (World Scientific, Singapore, 1991).
 
\bibitem{Horava}  P. Horava, Stability of Fermi surfaces and K theory,Ê
Ê{\it Phys. Rev. Lett.} {\bf 95}, 016405 (2005).


\bibitem{Mineev} G.E. Volovik and V.P. Mineev,  Investigation of 
singularities in superfluid $^3$He and liquid crystals  by homotopic
topology methods, 
{\it JETP} {\bf 45} 1186 - 1196 (1977); Particle like solitons in
superfluid $^3$He phases,  {\it JETP} {\bf 46}, 401--404  (1977).


\bibitem{Berloff} N.G. Berloff and P.H. Roberts, Motions in a bose condensate:
VI. Vortices in a nonlocal model, {\it J. Phys. A: Math. Gen.} {\bf 32}, 5611--5625 (1999).


\bibitem{Weinberg}  S. Weinberg, The cosmological constant
problem, {\it Rev. Mod. Phys.} {\bf  61}, 1 (1989).


\bibitem{Padmanabhan}
 T. Padmanabhan, Cosmological constant - the weight of the vacuum,
  {\it Phys. Rept.} {\bf 380},  235--320 (2003).


\bibitem{Myths} G.E. Volovik, Vacuum Energy: Myths and Reality, 
prepared for the special issue of {\it Int. J. Mod. Phys.} devoted to
dark energy and dark matter; gr-qc/0604062.


\bibitem{Eckstein} S. Eckstein and B.B. Varga, Dispersion of phonons  in
$^4$He, {\it  Phys. Rev. Lett.} {\bf 21}, 1311--1314 (1968).


\bibitem{Khalatnikov} I.M.  Khalatnikov, {\it An Introduction to the
Theory of Superfluidity}, (Benjamin, New York, 1965).
 
\bibitem{Fischer} U.R. Fischer, Dynamical aspects of analogue gravity:
The backreaction of quantum fluctuations in dilute Bose-Einstein
condensates, cond-mat/0512537.


\bibitem{SpecialRel} G.E. Volovik,  Vacuum energy and Universe in
special relativity,  {\it JETP Lett.} {\bf 77}, 639--641 (2003);
gr-qc/0304103.


\bibitem{Gusev-Zelnikov} Yu.V. Gusev and A.I. Zelnikov, 
Finite temperature nonlocal effective action for quantum fields in
curved space, {\it Phys. Rev.} {\bf D59}, 024002 (1998).

\bibitem{Volovik-Zelnikov} G.E. Volovik and A.I. Zelnikov, 
Universal temperature corrections to the free energy for the
gravitational field,  {\it JETP Lett.} {\bf 78}, 751--756 (2003);
gr-qc/0309066.  


\bibitem{Kirilin}   G. Kirilin and I. Khriplovich, Quantum power
correction to the Newton law, {\it JETP} {\bf 95},Ê 981--986 (2002).

\bibitem{Ilinski} K.N. Ilinski and A.S. Stepanenko, From Bose
condensation to quantum gravity and back, {\it J. Phys. Stud.} {\bf 2},
 155--159 (1998); cond-mat/9803233; Hydrodynamics of a Bose condensate: beyond the mean field
approximation (II), cond-mat/9612117;  First quantum corrections for a
hydrodynamics of a nonideal Bose gas, cond-mat/9607202.

\bibitem{Dobado} A. Dobado and A.L. Maroto, Particle production from
nonlocal gravitational effective action, {\it Phys. Rev.} {\bf D60},
104045 (1999).

\bibitem{Exotic} G.E. Volovik, {\it Exotic Properties of
Superfluid} $^3$He" (World Scientific, Singapore, 1992), Chapter 6.


\bibitem{Balbinot} R. Balbinot, S. Fagnocchi and A. Fabbri, 
Quantum effects in acoustic black holes: The backreaction,
{\it Phys. Rev.} {\bf D71}, 064019 (2005).


\bibitem{TunnelingVolovik} G. E. Volovik, Simulation of Painleve-Gullstrand black hole in
thin $^3$He-A film,  
{\it JETP Lett.} {\bf 69}, 705--713 (1999); gr-qc/9901077.

\bibitem{Parikh} M.K. Parikh and F. Wilczek, Hawking radiation as tunneling,
{\it Phys. Rev. Lett.} {\bf 85}, 5042--5045 (2000).

 \bibitem{Bjorken} J.D. Bjorken, private communications; 
J.D. Bjorken, Cosmology and the standard model,  
{\it Phys. Rev.} {\bf D67}, 043508 (2003).

\bibitem{KlinkhamerVolovik} F.R. Klinkhamer and G.E. Volovik, Merging
gauge coupling constants without Grand Unification, {\it JETP Lett.} {\bf
81},   551--555(2005); hep-ph/0505033.

  
\end{thebibliography}
\end{document}